%% file: Main.tex
\documentclass[lettersize,journal]{IEEEtran}
\usepackage{amsmath,amsfonts}
\usepackage{amsthm,amssymb}
\usepackage{mathrsfs,mathtools,mathabx}
\usepackage{bm}
\usepackage{algorithmic}
\usepackage{array}
\usepackage[caption=false,font=normalsize,labelfont=sf,textfont=sf]{subfig}
\usepackage{textcomp}
\usepackage{stfloats}
\usepackage{url}
\usepackage{verbatim}
\usepackage{booktabs}
\usepackage[table]{xcolor}
\usepackage{graphicx}
\def\BibTeX{{\rm B\kern-.05em{\sc i\kern-.025em b}\kern-.08em
    T\kern-.1667em\lower.7ex\hbox{E}\kern-.125emX}}
\usepackage{balance}
\usepackage{mycolors}
\usepackage{mytikzsetup}
\usepackage{fontawesome5}
\usepackage{tikz}
\usetikzlibrary{
    arrows.meta,
    positioning,
    calc,
    shapes.geometric
}

\begin{document}

\title{The Computing Channel: 

How Modulation Programs the Airwaves}
\author{ Saeed Razavikia,~\IEEEmembership{Member,~IEEE}, Carlo Fischione,~\IEEEmembership{Fellow,~IEEE}
\thanks{S. Razavikia and C. Fischione are with the School of Electrical Engineering and Computer Science, KTH Royal Institute of Technology, Stockholm, Sweden (e-mail: \{sraz, carlofi\}@kth.se). C. Fischione is also with Digital Futures of KTH. 

}
}

\maketitle

\begin{abstract}
Distributed computing and distributed artificial intelligence require frequent exchanges of intermediate results, although many applications need only an aggregate rather than individual device messages. Conventional systems recover each message before computing the aggregate, whereas over-the-air computation (OAC) exploits simultaneous transmission to obtain it directly. However, dominant OAC implementations rely on analog signaling, creating a mismatch with finite-precision data and digital communication procedures. This article presents digital function-oriented communication, in which finite-alphabet symbol representations and receiver decisions are jointly designed so that multiple-access superposition encodes the desired function without recovering individual inputs. We introduce its computational-constellation principle, main design approaches, extensions, and implementation challenges. Federated edge learning illustrates how the framework can reduce user-dependent data-bearing resources while operating directly on quantized model updates.
\end{abstract}


\section{Introduction}
\label{sec:channel-computing-platform}
\IEEEPARstart{D}{istributed} computing and distributed artificial intelligence are changing how data-intensive tasks are executed across cloud servers, edge devices, and wireless sensors. Instead of collecting all data at a central processor, these systems divide computation among multiple nodes. Local processing, however, does not eliminate communication. The participating nodes must repeatedly exchange intermediate results, including gradients, model updates, sensor measurements, and optimization variables. As the number of nodes and the frequency of these exchanges increase, data movement can become a major contributor to latency, energy consumption, and network-resource usage~\cite{Zhu2020IntelligentEdge,sapio2021scaling}.

Importantly, many of these exchanges are computations disguised as communication. A federated-learning server usually requires an aggregate of the local model updates rather than every update individually. Similarly, distributed sensing, consensus, and distributed optimization commonly require functions such as a sum, average, maximum, or majority decision. Conventional communication systems nevertheless encode the local values as user-specific messages, allocate communication resources to their delivery, recover the individual messages at the receiver, and only then evaluate the desired function. This architecture is necessary when the individual data must be reconstructed, but it transfers more information than the application ultimately consumes when only an aggregate is required.

Compression, sparsification, and improved multiple-access techniques can reduce this overhead. Nevertheless, they generally preserve the separation between communication and computation: the network transports information, and a processor subsequently computes the requested function. Over-the-air computation (OAC) provides a more direct physical-layer approach. It coordinates simultaneous transmissions over a shared communication resource and exploits waveform superposition to form an aggregate directly at the receiver~\cite{nazer2007Computation,Sahin2023AirCompSurvey}. Under suitable channel and synchronization conditions, one scalar aggregate can therefore be obtained without assigning a separate data-bearing resource to every participating node. OAC consequently changes interference from an impairment that must be removed into a computational operation that can be exploited.

The dominant realization of OAC has been analog. Local values are represented through continuous signal amplitudes so that their physical superposition approximates the desired aggregate. This approach can be highly communication efficient, but it creates an important implementation mismatch~\cite{golden2013harnessing}. Distributed computing platforms generate, store, and process data in finite-precision digital form, while contemporary communication systems are organized around finite-alphabet modulation, packetization, channel coding, and digital receiver processing. Here, \emph{digital} refers to the representation and processing of information; the transmitted radio waveform remains physically analog. Conventional analog OAC does not directly inherit these digital-link mechanisms and can be sensitive to noise, synchronization errors, channel misalignment, hardware impairments, and receiver dynamic-range limitations.

The central challenge is therefore to retain the concurrent-access advantage of OAC while operating directly on quantized data through finite-alphabet signaling. This objective aligns with the broader semantic and task-oriented communication paradigm, which designs communication according to the receiver’s intended task rather than requiring exact recovery of every source message~\cite{gunduz2022beyond}. It motivates digital function-oriented communication, in which the transmitter mappings and receiver decisions are jointly designed so that the superposed symbols represent the requested function. Each device maps its quantized input to a digital modulation symbol and transmits concurrently, while the receiver interprets the resulting computational constellation directly as a function output without necessarily recovering the individual inputs.

These limitations motivate the pursuit of a native digital solution for over-the-air computation. This paradigm shift raises a fundamental, function-oriented question:

\begin{center}
\setlength{\fboxsep}{7pt}
\colorbox{gray!7}{%
    \parbox{0.88\linewidth}{%
        \centering
     \emph{Can a multiple-access channel compute functions of distributed digital data while retaining the communication efficiency of over-the-air aggregation?} 
    }%
}
\end{center}

This question motivates a unified function-oriented constellation design in which the target function determines how digital symbols are represented. Depending on the design constraints, either the positions of the constellation points or their associated labels are tailored to the desired function. In both cases, the received superpositions are structured so that different input combinations producing the same function value need not be distinguished, whereas those yielding different values remain reliably distinguishable. The channel thereby becomes a programmable interface for computing functions of distributed digital data.

Fig.~\ref{fig:channel-evolution} illustrates the transition from individual-message recovery to digital function-oriented reception. This tutorial presents the fundamental principles of this paradigm, together with its main design approaches, extensions, and practical implementation challenges. Federated edge learning is considered as a representative application to demonstrate its potential for reducing user-dependent communication resources while retaining a digital representation of the exchanged model updates.

\input{figures/Fig_ChannelEvolution}

\section{Computing Through Function-Oriented Design}
\label{sec:computing-through-modulation}

To understand how digital modulation can directly represent a target computation, it is helpful to look at a simple scenario. Modulation can be designed not only to convey bit labels reliably but also to embody the computation itself. The following two-node BPSK example shows how a single induced constellation at the receiver can represent two completely different functions simply through different receiver labelings~\cite{Razavikia2024ChannelComp}.

\begin{center}
\setlength{\fboxsep}{6pt}
\colorbox{gray!7}{%
\parbox{0.91\columnwidth}{%

\smallskip
\centering
\textbf{BPSK example:} Each node maps its binary input $s_k \in \{0, 1\}$ to a physical signal $x_k \in \{-1, +1\}$, and the noiseless channel produces \(\rho=x_1+x_2\).

\medskip
\begin{tikzpicture}[x=1.02cm,y=0.48cm,font=\scriptsize]
    \draw[-{Latex[length=1.8mm]}] (-2.65,0) -- (2.65,0) node[right] {\(\rho\)};
    \foreach \x in {-2,0,2}{
        \fill (\x,0) circle (1.8pt);
        \node[above=2pt] at (\x,0) {\(\x\)};
    }

    \node[above=16pt] at (-2,0) {\((0,0)\)};
    \node[above=16pt,align=center] at (0,0) {\((0,1)\), \((1,0)\)};
    \node[above=16pt] at (2,0) {\((1,1)\)};

    \node[anchor=east,color={rgb, 255:red, 15; green, 83; blue, 26 } ] at (-2.65,-0.9) {$\mathscr{D}_{1}: {\color{black}{\rho}}  \mapsto s_{1} +s_{2}$};
    \node[color={rgb, 255:red, 15; green, 83; blue, 26 }] at (-2,-0.9) {\(0\)};
    \node[color={rgb, 255:red, 15; green, 83; blue, 26 }] at (0,-0.9) {\(1\)};
    \node[color={rgb, 255:red, 15; green, 83; blue, 26 }] at (2,-0.9) {\(2\)};

    \node[anchor=east,color={rgb, 255:red, 8; green, 80; blue, 177 } ] at (-2.65,-1.7) {$\mathscr{D}_{2}: {\color{black}{\rho}} \mapsto  s_{1} s_{2}$};
    \node[color={rgb, 255:red, 8; green, 80; blue, 177 } ] at (-2,-1.7) {\(0\)};
    \node[color={rgb, 255:red, 8; green, 80; blue, 177 } ] at (0,-1.7) {\(0\)};
    \node[color={rgb, 255:red, 8; green, 80; blue, 177 } ] at (2,-1.7) {\(1\)};
\end{tikzpicture}

\smallskip
\raggedright
}%
}
\end{center}

More generally, consider \(K\) nodes connected to a common receiver. Node $k$ observes a finite-alphabet digital value $s_k$ and maps it directly to a channel input $x_k$. When the nodes access the channel concurrently, computation and communication can be represented by
\begin{equation}
    \rho \sim p_{\rho|x_1,\ldots,x_K}
    \bigl(\,\cdot\,|x_1,\ldots,x_K\bigr),
    \qquad
    \widehat{f}=\mathscr{D}(\rho),
    \label{eq:channelcomp-model}
\end{equation}
where \(p_{\rho|x_1,\ldots,x_K}\) denotes the multiple-access channel law and \(\rho\) is the receiver observation. For a linear additive channel, this law is induced by a weighted superposition of the channel inputs, where the noisy is commonly modeled as additive input-independent Gaussian noise. The channel-law formulation also covers shared media for which an additive-noise representation is inappropriate. In a direct-detection optical multiple-access channel, for example, the users transmit nonnegative optical intensities, and the receiver observes a photon count whose conditional distribution is Poisson. Its mean depends on the combined received intensity and the background radiation; hence, the uncertainty is inherently signal dependent. The Poisson multiple-access channel has been established as a model for many-to-one optical communication through fiber or free space~\cite{lapidoth1998poisson}. More generally, computation over multiple-access channel laws is studied in~\cite{nazer2007Computation}. Accordingly, the programming principle is governed by the aggregation law of the physical medium, while the signaling mappings and receiver must be designed jointly for the target function and the corresponding channel statistics.


\input{figures/Fig_computationalmodulation}


Conventional digital links recover individually labeled messages before performing the required computation. In contrast, ChannelComp jointly designs the transmitter mappings and receiver rule to estimate the desired function directly from the aggregate channel response, without necessarily recovering the individual source values. Each input tuple induces a representation at the receiver, and the collection of all such representations forms the \emph{induced computational constellation}. For an additive channel, a representation corresponds to a nominal received value; for a stochastic channel, it may correspond to a parameter of the conditional output distribution, such as the mean photon-count rate of a Poisson channel.

Formally, for an input tuple $\mathbf{s}=(s_1,\ldots,s_K)$, let $f(\mathbf{s})$ denote the desired function value and $\rho(\mathbf{s})$ its induced channel representation. 
Two input tuples may share the same representation when they yield the same function value. A collision is destructive only when different function values become indistinguishable. Therefore, exact computation under the nominal model requires
\begin{align}
    \rho(\mathbf{s})\neq\rho(\mathbf{s}')
\quad\text{whenever}\quad
f(\mathbf{s})\neq f(\mathbf{s}').
\end{align}
Under noise, reliability depends on the distinguishability of the corresponding conditional channel distributions. Importantly, \(\rho(\mathbf{s})\) need not equal \(f(\mathbf{s})\); it only needs to identify the desired function value. Quantization separately determines the discrepancy from the function of the original continuous observations.

While function-oriented mappings can structurally prevent destructive collisions, collision-free mapping alone does not guarantee reliable computation under noise or channel mismatch. The critical metric is the decision margin—or, more generally, the channel-dependent distinguishability—between representations associated with different outputs. This margin must be sufficiently large relative to noise, channel-estimation errors, and other physical modeling uncertainties.

To achieve this, most of the design optimization is performed offline using the target function, input alphabets, channel model, active-node configuration, and resource constraints. During online operation, each node simply quantizes its observation, selects the corresponding precomputed symbol, and transmits concurrently. The receiver then identifies the resulting function value using the predetermined decision rule. Although changes in the active-node set or substantial departures from the assumed channel model may require receiver adaptation, robust design, or recomputation of the mappings, this framework fundamentally shifts the heavy design burden to offline system configuration, retaining simple and low-latency online operation under the modeled conditions~\cite{Razavikia2024ChannelComp}.

\input{figures/Fig_SumComp}

\section{Reliable and Scalable Computation}
\label{sec:reliable-scalable-modulation}

Collision-free induced constellations establish noiseless feasibility but do not ensure reliable computation under random disturbances and channel uncertainty. Moreover, exhaustive offline design becomes prohibitive as the number of nodes increases, and many applications require vector-valued rather than scalar outputs. These challenges motivate three complementary mechanisms: task- and distribution-aware geometry for reliability, symmetry-based reduction for scalability, and spatial processing for parallel computation.

\subsection{Distribution-Aware Design }

Conventional modulation generally protects distinct symbol labels using a common geometric criterion. In computational modulation, however, not all decision errors have the same consequence. Confusing two outputs that are similar for the target application may incur only a small loss, whereas confusing outputs with substantially different meanings can be more damaging. The induced geometry should therefore account for both the statistical distinguishability of the channel observations and the functional consequence of an incorrect decision.

Using the notation introduced in the preceding section, a compact design principle is
\begin{equation}
    \mathcal{D}_{\mathrm{ch}}(\rho(\mathbf{s}),\rho(\mathbf{s}'))
    \geq
    \lambda\,\mathcal{L}\big(f(\mathbf{s}),f(\mathbf{s}')\big),
    \qquad f(\mathbf{s})\neq f(\mathbf{s}'),
    \label{eq:weighted-separation}
\end{equation}
where \(\mathcal{D}_{\mathrm{ch}}\) evaluates the channel-dependent distinguishability between the observations induced by \(\rho(\mathbf{s})\) and \(\rho(\mathbf{s}')\), e.g., an additive channel with heavy-tailed noise, this might be measured by the squared Euclidean separation between the induced representations, whereas likelihood-based divergences are better suited for general channel laws. Also  \(\mathcal{L}(f(\mathbf{s}),f(\mathbf{s}'))\) quantifies the loss incurred by confusing the corresponding outputs. For scalar outputs, this is typically calculated as the squared difference between the function values. The transmitter mappings are designed to maximize the smallest separation margin  \(\lambda\) across the relevant input pairs.

Consequently, no single induced geometry is optimal for every operating condition. The preferred design depends jointly on the signal-to-noise ratio, channel statistics, input probabilities, and application loss. At high signal-to-noise ratios, many collision-free geometries may provide adequate reliability. Under stronger or non-Gaussian disturbances, explicitly distribution- and task-aware designs can provide greater protection for the computational decisions that matter most.

\subsection{Beamforming for Vector Computation}

Distributed learning and sensing commonly require an \(L\)-dimensional vector output, \(\mathbf{f}=[f_1,\ldots,f_L]^{\mathsf T}\). To preserve the latency advantages of concurrent transmission, these \(L\) components can be computed simultaneously by exploiting parallel spatial modes via multiple antennas. 

In a MIMO implementation, each node applies a transmit beamforming matrix, while the receiver employs a combining matrix. Unlike conventional multiuser MIMO, which spatially separates individual data streams, computational beamforming purposefully aligns device contributions associated with the same output. This preserves the desired aggregate while limiting interference between the \(L\) parallel computational streams.

Crucially, this vector extension does not require instantaneous CSI at the distributed transmitters. As shown in~\cite{Razavikia2026VecComp}, the transmit matrices can be generated statistically using only large-scale channel information for power control, while fast channel adaptation is performed at the receiver. This division reduces CSI feedback, channel inversion, and online processing at energy-constrained nodes; thus, scalability here refers primarily to device-side overhead rather than antenna resources.

\input{figures/Table_source}

\subsection{Pyramid Sampling for Scalability}

The offline design of ChannelComp becomes increasingly complex as the number of nodes and quantization levels grows. A direct design must examine a rapidly expanding set of input combinations and ensure adequate separation between combinations producing different function outputs. For symmetric functions with a common encoder across the nodes, however, permutations of the same quantized values produce identical function outputs and received sums. These redundant combinations can therefore be grouped according to how frequently each quantization level occurs. Pyramid sampling further reduces the design burden by retaining only a structured subset of these groups. Its sampling order determines the density of the retained cases and thereby controls the trade-off between design complexity and approximation accuracy~\cite{razavikia2025designing2}.

The effect of this reduction depends on the target function. For the examined sampling orders of one, two, four, and eight, the optimized constellation for summation preserves its PAM-like geometry. The product function is more sensitive: two constellation points gradually approach each other as the sampling becomes coarser and eventually coincide at the highest tested order. The sampling order must therefore be selected according to the function and the required computational accuracy.

A more aggressive approximation restricts the design to selected trajectories through the sampled input space. In the considered eight-level example, this approach reduces the design to only 28 representative comparisons. However, it can omit relevant pairs outside those trajectories and may leave some conflicting outputs insufficiently separated. A practical solution is to inspect all retained cases after the initial design and iteratively add any violated comparisons. The final design should also be evaluated over the complete input space to determine the error introduced by sampling.

A separate ideal-channel experiment with five nodes reported an MSE reduction of up to two orders of magnitude for majority computation. In that comparison, a higher sampling order was combined with a substantially finer quantization grid. The result therefore reflects a joint complexity--accuracy trade-off enabled by coarser sampling and finer quantization; it should not be interpreted as an accuracy improvement caused by pyramid sampling alone~\cite{razavikia2025designing2}.

\input{figures/Fig_fed}

\subsection{Computing with QAM Constellations}

To compute functions over standard constellations without designing new geometries from scratch, we can employ a structured approach called the SumComp scheme~\cite{Razavikia2025SumComp}. For summation tasks, SumComp retains prescribed PAM, QPSK, or rectangular-QAM grids, but replaces conventional Gray labeling with an arithmetic-preserving mapping. Quantized integers are assigned to available signal points such that their over-the-air superposition has an unambiguous arithmetic interpretation. This allows the receiver to directly recover the aggregate sum without needing to detect individual device inputs.

Because the physical symbol coordinates and RF chains remain unchanged, the proposed scheme effectively reuses standard modulation hardware. It only requires a custom digital mapper at the transmitter and an aggregate-aware detector at the receiver. Figure~\ref{fig:Hexagonal} illustrates this flexibility using an order-8 hexagonal-QAM constellation. Here, the SumComp coding parameters dictate how integer values are distributed across the lattice, cleanly altering the representable input range while guaranteeing that addition is preserved under superposition.

When multiple communication resources are available, the scheme's reliability can be further enhanced through bit slicing~\cite{liu2025digital}. This technique divides a high-precision quantized value into shorter segments, transmitting them across several lower-order symbols. By replacing one dense constellation with multiple sparser ones, the system exchanges extra channel uses for stronger protection of the most significant bits.

Finally, because the SumComp scheme inherently enlarges the aggregate constellation as the number of active nodes grows, conventional single-user detection cannot be applied directly. The receiver requires adequate dynamic range and ADC resolution, along with precise power, timing, and phase alignment. Provided these conditions are met, the proposed scheme seamlessly enables in-network summation over a shared channel using standard finite-alphabet signals.

\input{figures/Table_compare}

\section{Use Case: Federated Edge Learning}

Federated edge learning is a direct application of channel programming. During each learning round, the edge server requires an aggregate of the local gradients or model updates. A conventional communication system nevertheless delivers and decodes the update from every device before calculating this aggregate. Earlier work showed that wireless superposition can perform the aggregation directly using compressed analog transmission, gradient sparsification, dimensionality reduction, and channel-aware power allocation~\cite{amiri2020federated}. As a digital alternative, the implementation in~\cite{razavikia2024blind} uses function-oriented QAM signaling. Each device quantizes its local update and maps every coordinate to a QAM symbol. The devices transmit the symbols corresponding to the same coordinate concurrently, and their superposition represents the required sum. The server then obtains the average through aggregate-aware decoding and appropriate scaling. Because this procedure operates on the quantized outputs of local training, it remains independent of the learning algorithm and can support different optimizers and model architectures.

A shared transmission processes one coordinate of the model update rather than the entire update vector. All participating devices transmit their symbols for that coordinate over the same data-bearing resource element, and the procedure is repeated for the remaining coordinates. When several OFDM subchannels are available, multiple coordinates can be transmitted in parallel. Consequently, the required number of transmission intervals depends on the number of transmitted coordinates and the available parallel subchannels, but it does not increase with the number of participating devices.

Table~\ref{tab:feel-resource-accounting} clarifies this resource advantage. Under the stated assumptions, an orthogonal digital baseline allocates separate data-bearing resources to every device and every transmitted coordinate. By contrast, the aggregation schemes allow all devices to share the same resource for each coordinate. They therefore remove the device-count factor from the data-bearing resource requirement. This comparison does not imply a universal reduction in latency or bandwidth. Compression and sparsification can reduce the number of transmitted coordinates, while channel coding, modulation, pilot transmission, protocol overhead, and multiuser MIMO can alter the overall resource usage. Thus, the QAM realization avoids resource growth with the number of devices under the adopted assumptions, although its resource requirement still increases with the model-update dimension.

Fading introduces a second programming challenge: the arithmetic relationship embedded in the transmitted symbols must remain observable after propagation. Analog federated-learning schemes commonly use transmitter-side channel knowledge and channel-aware power allocation to align device contributions at the server~\cite{amiri2020federated}. The finite-alphabet realization in~\cite{razavikia2024blind} instead shifts instantaneous channel compensation to the edge server. The devices transmit without instantaneous channel state information, while the server estimates their aggregate channel and exploits spatial averaging across its receive antennas. Under independent Rayleigh fading, equal channel variances, accurate aggregate-channel estimation, and perfect synchronization, the resulting fading error decreases approximately inversely with the number of receive antennas. Independent experimental studies illustrate both the feasibility and the limitations of these assumptions. For example, a digital OAC implementation combined OFDM with convolutional or LDPC coding and joint aggregation decoding, demonstrating real-time operation with multiple USRP transmitters under phase asynchrony~\cite{you2023broadband}. Such prototype results establish the feasibility of sum aggregation in calibrated experimental configurations; however, they do not by themselves resolve active-user coordination, arbitrary finite-alphabet function programming, or standards-compatible reliability procedures.

\section{Outlook}
\label{sec:outlook}

Programmable channels shift part of a distributed application's workload from the processor directly into the communication medium. Inspired by the success of network-layer aggregation frameworks like SwitchML~\cite{sapio2021scaling}, this principle can be extended to the physical layer. By jointly designing modulation and superposition, the wireless channel directly computes a low-dimensional aggregate from distributed inputs, rather than merely transporting individual messages for later computation. However, transitioning from isolated modulation constructions to broad deployment requires addressing fundamental, interlocking challenges in scalability, physical-layer reliability, and network security.

End-to-end constellation design must be executed jointly with quantization, multidimensional encoding mappings, and the dynamic allocation of spatial and temporal resources. As network dimensions increase, the prohibitive complexity of offline optimization—even when mitigated by techniques like pyramid sampling—demands adaptive, learning-assisted optimization and tractable design criteria that account for complex channel laws. Simultaneously, practical deployments must manage physical non-idealities by exploring pre-equalization strategies that leverage partial or full CSI to ease receiver complexity. Moving forward, the field requires a unified reliability framework that seamlessly integrates basic aggregate decoding~\cite{you2023broadband} with advanced forward error correction via non-binary linear channel codes, incremental redundancy, and dynamic link adaptation.

Finally, security and privacy must be established as explicit physical-layer design objectives. Recovering only an aggregate is not equivalent to cryptographic secure aggregation and does not automatically provide differential privacy; small participant groups or auxiliary information may still enable the inference of individual updates. Furthermore, malicious devices can easily poison the physical-layer sum by transmitting manipulated updates or noncompliant waveforms. Robust systems therefore require authenticated participation, waveform-integrity checks, and bounded contributions, all designed jointly with physical-layer reliability. By addressing these challenges, the approaches summarized in Table~\ref{tab:method-synthesis} can evolve into adaptive, verifiable, and highly secure function-native communication systems.

\bibliographystyle{IEEEtran}
\bibliography{references}



\end{document}

%% file: figures/Fig_ChannelEvolution.tex
\begin{figure*}
    \centering
 \subfloat[Recover all messages first]{ 

    \tikzset{every picture/.style={line width=0.75pt}} 
    \scalebox{0.55}{
\begin{tikzpicture}[x=1cm,y=1cm]

\node[
    anchor=west,
    text=mainblue,
    font=\bfseries\Large
] at (3.2,6)
{Conventional communication channel};

\node[nodebox] (n1) at (2,4.00)
    {\textbf{Node $1$}\\[-1mm] $s_1$};

\node[nodebox] (nk) at (2,2.65)
    {\textbf{Node $2$}\\[-1mm] $s_2$};

\node[nodebox] (nK) at (2,0.85)
    {\textbf{Node $K$}\\[-1mm] $s_K$};

\node[font=\Huge] at (2,1.78) {$\vdots$};

\node[channelbox] (channel) at (7.05,2.65) {};

\node[
    align=center,
    text=mainblue,
    font=\bfseries\small
] at (7.05,3.65)
{Channel\\as data pipe};

\node[
    cylinder,
    draw=mainblue,
    line width=1.1pt,
    shape border rotate=180,
    aspect=1.6,              
    minimum height=1.7cm,     
    minimum width=0.76cm,     
    cylinder uses custom fill,
    cylinder body fill=blue!8,
    cylinder end fill=blue!15
] (pipe) at (7.07,2.10) {};

\draw[
    flowarrow, 
    line width=1.05pt
] ([xshift=-0.35cm]pipe.center) -- ([xshift=0.45cm]pipe.center);

\draw[flowarrow, color=bluegray]
    (n1.east) -- (channel.west |- n1.east);

\node[
    align=center,
    font=\normalsize
] at (4.10,4.08)
{slot / band 1\\[1mm]\small (orthogonal)};

\draw[flowarrow, color=burntsienna]
    (nk.east) -- (channel.west |- nk.east);

\node[
    align=center,
    font=\normalsize
] at (4.10,2.73)
{slot / band $2$\\[1mm]\small (orthogonal)};

\draw[flowarrow, color=cadmiumgreen]
    (nK.east) -- (channel.west |- nK.east);

\node[
    align=center,
    font=\normalsize
] at (4.10,0.93)
{slot / band $K$\\[1mm]\small (orthogonal)};

\node[inddecoderbox] (decoder) at (10.88,2.65) {};

\node[
    align=center,
    font=\small
] at (10.88,3.72)
{\textbf{Individual}\\\textbf{decoding}};

\node[
    align=center,font= \large,
] at (10.88,2.15)
{$\hat{s}_1 = \mathscr{D}(\rho_1)$\\$\hat{s}_2=\mathscr{D}( \rho_2)$\\[1mm]
 ${ \vdots}$\\[1mm]
 $\hat{s}_K= \mathscr{D}(\rho_K)$};

\draw[flowarrow]
    (channel.east) -- (decoder.west);

\node[computebox] (compute) at (14.25,2.65) {};

\node[
    align=center,
    font=\small
] at (14.25,3.72)
{\textbf{Compute}\\[1mm]
  $ f(\hat{s}_1,\ldots,\hat{s}_K)$};

\draw[flowarrow]
    (decoder.east) -- (compute.west);

\draw[
    mainblue,
    line width=1.15pt,
    rounded corners=1pt
] (13.82,1.52) rectangle (14.68,2.36);

\draw[
    mainblue,
    line width=1.15pt,
    rounded corners=1pt
] (13.98,1.68) rectangle (14.52,2.20);

\foreach \y in {1.64,1.84,2.04,2.24}{
    \draw[mainblue,line width=1pt]
        (13.60,\y) -- (13.82,\y);
    \draw[mainblue,line width=1pt]
        (14.68,\y) -- (14.90,\y);
}

\foreach \x in {13.94,14.14,14.34,14.54}{
    \draw[mainblue,line width=1pt]
        (\x,2.36) -- (\x,2.58);
    \draw[mainblue,line width=1pt]
        (\x,1.30) -- (\x,1.52);
}


\end{tikzpicture}}

}
 \subfloat[The channel naturally computes the sum of the signals]{   
    \scalebox{0.55}{
    \begin{tikzpicture}[x=1cm,y=1cm]

\node[
    anchor=west,
    text=mainblue,
    font=\bfseries\Large
] at (3.2,5.55)
{ Channel as signal superposition medium };
\node[nodebox] (n1) at (2.55,3.65)
    {\textbf{Node 1}\\[-1mm] $s_1$};

\node[nodebox] (nk) at (2.55,2.25)
    {\textbf{Node $2$}\\[-1mm] $s_2$};

\node[nodebox] (nK) at (2.55,0.45)
    {\textbf{Node $K$}\\[-1mm] $s_K$};

\node[font=\Huge] at (2.55,1.48) {$\vdots$};

\coordinate (a1) at (5.72,3.15);
\coordinate (ak) at (5.72,2.35);
\coordinate (aK) at (5.72,1.82);

\draw[flowarrow,color=scatterpurple] (n1.east) -- (a1);
\draw[flowarrow,color=scatterpurple] (nk.east) -- (ak);
\draw[flowarrow,color=scatterpurple] (nK.east) -- (aK);

\node at (3.9,3.72) {$x_1$};
\node at (3.9,2.62) {$x_2$};
\node at (3.9,1.65) {$x_K$};

\path (n1.east) -- (a1) coordinate[pos=0.52] (m1);
\path (nk.east) -- (ak) coordinate[pos=0.52] (mk);
\path (nK.east) -- (aK) coordinate[pos=0.52] (mK);

\begin{scope}[shift={($(m1)+(0,0.36)$)}, x=0.24cm, y=0.14cm]
    \draw[scatterpurple!85!black, line width=0.8pt]
    plot[smooth,domain=-1.6:1.6,samples=80]
    (\x,{0.70*sin(360*\x)});
\end{scope}

\begin{scope}[shift={($(mk)+(0,0.36)$)}, x=0.24cm, y=0.14cm]
    \draw[scatterpurple!85!black, line width=0.8pt]
    plot[smooth,domain=-1.6:1.6,samples=80]
    (\x,{0.55*cos(300*\x)});
\end{scope}

\begin{scope}[shift={($(mK)+(0,0.36)$)}, x=0.24cm, y=0.14cm]
    \draw[scatterpurple!85!black, line width=0.8pt]
    plot[smooth,domain=-1.6:1.6,samples=80]
    (\x,{0.38*sin(420*\x) + 0.22*cos(240*\x)});
\end{scope}

\begin{scope}[scale=0.04, shift={(40cm,155cm)}, yscale=-1]
  \fill[color=black]  (125,85) circle  (2.5cm);

  \draw[thick, color=bazaar!60!black!80] (137,75) arc[start angle=-30, end angle=30, radius=18];
  \draw[thick, color=bazaar!60!black!80] (132,80) arc[start angle=-30, end angle=30, radius=10];

  \draw[thick, color=bazaar!60!black!80] (113,93) arc[start angle=150, end angle=210, radius=18];
  \draw[thick, color=bazaar!60!black!80] (118,90) arc[start angle=150, end angle=210, radius=10];

  \draw[very thick] (110,135) -- (125,90);
  \draw[very thick] (140,135) -- (125,90);

  \draw[thick] (112,128) -- (133,115);
  \draw[thick] (138,128) -- (117,115);

  \draw[thick] (133,115) -- (120,105);
  \draw[thick] (117,115) -- (130,105);

  \draw[thick] (120,105) -- (128,98);
  \draw[thick] (130,105) -- (122,98);
\end{scope}

\draw[flowarrow] (7.28,2.65) -- (8.40,2.65);

\draw[flowarrow] (11.8,2.65) -- (13,2.65);


\node[sumdecoderbox] (fblock) at (10.2,2.65) {
    \textbf{Decode the sum} \\[3mm] 
    {\Large$\hat{a} := \mathscr{D}(\rho)$} \\[3mm]
    {\normalsize where $a = \sum_k s_k$}
};


\node[funcbox] (fblock2) at (15.5,2.65)
{
    \textbf{Compute by post processing}\\[1mm]
    {\Large $ \widehat{f} := \psi(\hat{a})$}
};

\end{tikzpicture}}}

\subfloat[Design transmitted symbols so that superposition represents the desired function]{

\scalebox{0.65}{
\begin{tikzpicture}[
    x=1cm,
    y=1cm,
    every node/.style={font=\small}
]

\node[
    anchor=west,
    text=mainblue,
    font=\bfseries\Large
] at (5.2,6)
{Channel as computation  medium};

\node[sourcebox] (n1) at (-1,4.55)
    {\textbf{Node 1}\\[-1mm] $s_1$};

\node[sourcebox] (n2) at (-1,2.85)
    {\textbf{Node 2}\\[-1mm] $s_2$};

\node[sourcebox] (nK) at (-1,0)
    {\textbf{Node $K$}\\[-1mm] $s_K$};

\node[font= \Huge] at (0.10,1.52) {$\vdots$};

\node[encoderbox] (e1) at (1.50,4.55)
    {\Large $\mathscr{E}_1$};

\node[encoderbox] (e2) at (1.50,2.85)
    {\Large  $\mathscr{E}_2$};

\node[encoderbox] (eK) at (1.50,0)
    {\Large  $\mathscr{E}_K$};

\node[font=\Huge] at (2.8,1.50) {$\vdots$};

\draw[flowarrow]
    (n1.east) -- (e1.west);

\draw[flowarrow]
    (n2.east) -- (e2.west);

\draw[flowarrow]
    (nK.east) -- (eK.west);


\localconstellation
    {4.8}{4.55}
    {(0,0)}
    {bluepoint}

\localconstellation
    {4.8}{2.55}
    {(0.26,0)}
    {orangepoint}

\localconstellation
    {4.8}{0}
    {(-0.26,0)}
    {greenpoint}


\coordinate (m1) at (6.10,3.62);
\coordinate (mk) at (6.10,3.05);
\coordinate (mK) at (6.10,2.17);

\draw[flowarrow]
    (e1.east) -- (3.3,4.55);

\draw[flowarrow]
    (e2.east) -- (3.3,2.85);

\draw[flowarrow]
    (eK.east) -- (3.3,0);

\coordinate (a1) at (8.10,3.92);
\coordinate (ak) at (8.10,3.05);
\coordinate (aK) at (8.10,2.17);

\draw[flowarrow,color=scatterpurple]
    ([yshift=1.05cm]m1.east) -- ([yshift=0.5cm,xshift=.2cm] ak);

\draw[flowarrow,color=scatterpurple]
    (mk.east) -- ( [xshift=.2cm]ak);

\draw[flowarrow,color=scatterpurple]
    ([yshift=-1.55cm]mK.east) -- ([yshift=-0.4cm,xshift=.2cm] ak);

\node[above] at (6.63,3.72) {$x_1$};
\node[above] at (6.63,2.98) {$x_2$};
\node[above] at (6.63,1.55) {$x_K$};


\begin{scope}[scale=0.05, shift={(60cm,150cm)}, yscale=-1]
  \fill[color=black]  (125,85) circle  (2.5cm);

  \draw[thick, color=bazaar!60!black!80] (137,75) arc[start angle=-30, end angle=30, radius=18];
  \draw[thick, color=bazaar!60!black!80] (132,80) arc[start angle=-30, end angle=30, radius=10];

  \draw[thick, color=bazaar!60!black!80] (113,93) arc[start angle=150, end angle=210, radius=18];
  \draw[thick, color=bazaar!60!black!80] (118,90) arc[start angle=150, end angle=210, radius=10];

  \draw[very thick] (110,135) -- (125,90);
  \draw[very thick] (140,135) -- (125,90);

  \draw[thick] (112,128) -- (133,115);
  \draw[thick] (138,128) -- (117,115);

  \draw[thick] (133,115) -- (120,105);
  \draw[thick] (117,115) -- (130,105);

  \draw[thick] (120,105) -- (128,98);
  \draw[thick] (130,105) -- (122,98);
\end{scope}

\draw[flowarrow]
    (10.5,3) -- (11.5,3);

\begin{scope}[shift={(0.5cm,0)}]


\node[decoderbox] (decoder) at (12.85,2.78) {};

\node[
    align=center,
    text=mainblue,
    font=\small\bfseries
] at (12.85,4.23)
{Direct function decoding};


    \clip (11.67,1.58) rectangle (13.42,3.34);

    \pgfmathsetseed{18}

    \foreach \i in {1,...,550}{
        \pgfmathsetmacro{\rx}
            {(rnd+rnd+rnd-1.5)*0.73}

        \pgfmathsetmacro{\ry}
            {(rnd+rnd+rnd-1.5)*0.68}

        \fill[
            scatterpurple,
            opacity=0.48
        ]
        ({12.85+\rx},{2.35+\ry}) circle (0.018);
    }
\end{scope}

\begin{scope}[shift={(-2.8cm,0.1cm)}]

\draw[axisarrow]
    (14.95,2.15) -- (17.12,2.15);

\draw[axisarrow]
    (16.05,1.18) -- (16.05,3.58);

\node[right] at (17.08,2.15) {$I$};
\node[above] at (16.05,3.52) {$Q$};

\draw[
    dashed,
    mainblue,
    line width=0.65pt
]
    (16.05,2.15) -- (14.96,2.92);

\draw[
    dashed,
    mainblue,
    line width=0.65pt
]
    (16.05,2.15) -- (16.48,3.43);

\draw[
    dashed,
    mainblue,
    line width=0.65pt
]
    (16.05,2.15) -- (17.02,2.75);

\node[
    circle,
    draw=mainblue,
    fill=signalblue,
    minimum size=2.3mm,
    inner sep=0pt
] at (15.53,3.02) {};

\node[
    diamond,
    draw=orangeborder,
    fill=burntsienna,
    minimum size=3mm,
    inner sep=0pt
] at (16.77,2.83) {};

\node[
    rectangle,
    draw=cadmiumgreen,
    fill=cadmiumgreen,
    minimum width=2.8mm,
    minimum height=2.8mm,
    inner sep=0pt
] at (15.35,1.55) {};

\node[
    regular polygon,
    regular polygon sides=3,
    shape border rotate=0,
    draw=red!60!black,
    fill=symbolred,
    minimum size=3.2mm,
    inner sep=0pt
] at (16.78,1.46) {};

\end{scope}
\draw[flowarrow]
    ([yshift=0.15cm]  decoder.east) -- ++(1,0);

\begin{scope}[shift={(-2cm,0)}]
    \node[
    circle,
    draw=mainblue,
    fill=signalblue,
    minimum size=2.8mm,
    inner sep=0pt
] at (19.00,4.16) {};

\node[anchor=west] at (19.38,4.16) {$f_1$};

\node[
    diamond,
    draw=orangeborder,
    fill=burntsienna,
    minimum size=3mm,
    inner sep=0pt
] at (19.00,3.43) {};

\node[anchor=west] at (19.38,3.43) {$f_2$};

\node[
    rectangle,
    draw=cadmiumgreen,
    fill=cadmiumgreen,
    minimum width=2.6mm,
    minimum height=2.6mm,
    inner sep=0pt
] at (19.00,2.70) {};

\node[anchor=west] at (19.38,2.70) {$f_3$};

\node[
    regular polygon,
    regular polygon sides=3,
    shape border rotate=0,
    draw=red!60!black,
    fill=symbolred,
    minimum size=2.8mm,
    inner sep=0pt
] at (19.00,1.97) {};

\node[anchor=west] at (19.38,1.97) {$f_4$};

\node[font=\Huge] at (19.00,1.22) {$\vdots$};
\end{scope}


\end{tikzpicture}}
}

\caption{Evolution from conventional message-oriented communication to function-oriented computation over a shared channel. Conventional systems first recover the individual messages and then evaluate the target function, whereas function-oriented modulation maps simultaneous transmissions to an induced constellation whose decision regions represent the function values.}
    \label{fig:channel-evolution}
\end{figure*}
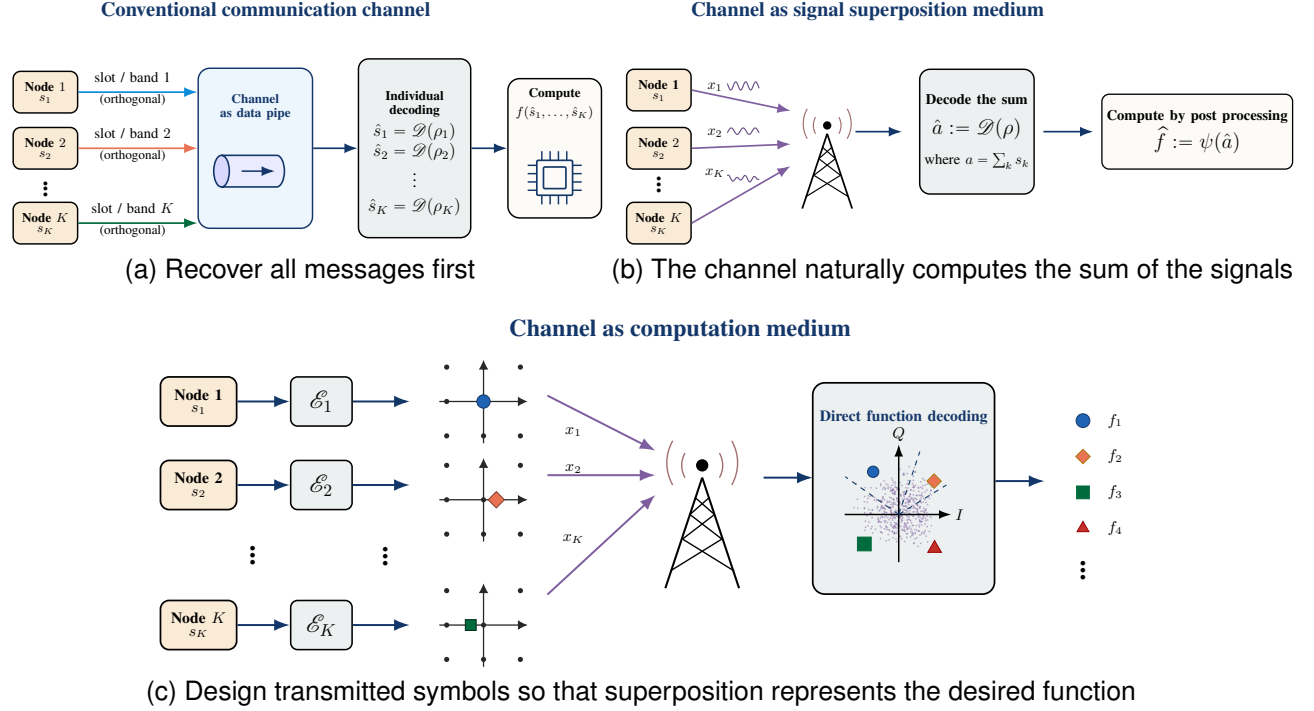

%% file: figures/Fig_computationalmodulation.tex
\begin{figure*}

\subfloat[ \footnotesize Distribution-aware geometry]{
    \centering
    \scalebox{0.7}{
    \begin{tikzpicture}[
    x=1cm,
    y=1cm,
    every node/.style={font=\small}
]

\node[
    anchor=west,
    font=\bfseries
] at (0.22,11.72)
{(a) Gaussian-like noise};

\begin{scope}[shift={(0.55,8.40)}]

    \draw[axis] (0,0) -- (2.15,0);
    \draw[axis] (0,0) -- (0,2.05);

    \node[anchor=north] at (0,0) {$0$};
    \node[anchor=north] at (2.05,0) {\Large  $\rho$};
    \node[anchor=east] at (0,1.92) {\Large  $p(\rho)$};

    \draw[
        mainblue,
        line width=1.15pt
    ]
    plot[
        domain=0:1.85,
        samples=120,
        smooth
    ]
    (
        {\x},
        {1.52*exp(-9.2*(\x-0.88)^2)}
    );

    \draw[
        dashed,
        black,
        line width=0.7pt
    ] (0.88,0) -- (0.88,1.55);

\end{scope}

\begin{scope}[shift={(5.18,9.05)}]

    \draw[boundary]
        (-2.10,-0.25)
        .. controls (-1.25,0.05) and (-0.30,0.30) ..
        (0.18,0.72);

    \draw[dashedboundary]
        (0.18,0.72)
        .. controls (0.72,1.5) and (0.86,1.70) ..
        (1,2);

    \draw[boundary]
        (0.18,0.72)
        .. controls (0.36,0.30) and (0.53,0.02) ..
        (0.86,-0.30);

    \draw[dashedboundary]
        (0.86,-0.30)
        .. controls (1.32,-0.18) and (1.77,0.02) ..
        (2.16,0.20);

    \draw[dashedboundary]
        (0.86,-0.30)
        .. controls (0.54,-0.82) and (0.48,-1.32) ..
        (0.45,-1.90);

    \gaussianhalo{(-1.30,1.30)}{mainblue}{0.92}
    \gaussianhalo{( 0.95,1.08)}{mainblue}{0.78}
    \gaussianhalo{(-0.15,1.68)}{mainblue}{0.78}

    \gaussianhalo{(-0.98,-1.05)}{halogreen}{1.12}

    \gaussianhalo{(-0.45,0.00)}{halogreen}{0.82}
    \gaussianhalo{( 0.95,-1.08)}{haloorange}{0.92}
    \gaussianhalo{( 1.82,-1.05)}{haloorange}{0.72}

    \node[bluepoint]  at (-0.15,1.68) {};
    \node[bluepoint]   at (-1.30,1.30) {};
    \node[bluepoint]   at ( 0.95,1.08) {};

    \node[greenpoint]  at (-0.98,-1.05) {};
    
     \node[greenpoint] at (-0.45,0.00) {};
    
    \node[orangepoint] at ( 0.95,-1.08) {};
    \node[orangepoint] at ( 1.82,-1.05) {};

\end{scope}

\draw[
    color=black,
    line width=0.8pt
] (0.45,7.30) -- (7.80,7.30);

\node[
    anchor=west,
    font=\bfseries
] at (0.22,6.72)
{(b) Heavy-tailed / impulsive noise};

\begin{scope}[shift={(0.8,3.55)}]

    \draw[axis] (0,0) -- (2.15,0);
    \draw[axis] (0,0) -- (0,2.05);

    \node[anchor=north] at (0,0) {$0$};
    \node[anchor=north] at (2.05,0) {\Large  $\rho$};
    \node[anchor=east] at (0,1.92) {\Large  $p(\rho)$};

    \draw[
        purple!75!black,
        line width=1.15pt
    ]
    plot[
        domain=0:1.95,
        samples=140,
        smooth
    ]
    (
        {\x},
        {1.74/(1+3.7*\x)^1.35}
    );

\end{scope}

\begin{scope}[shift={(5.18,4.22)}]

    \draw[boundary]
        (-2.10,-0.22)
        .. controls (-1.30,0.02) and (-0.38,0.32) ..
        (0.18,0.70);

    \draw[dashedboundary]
        (0.18,0.70)
        .. controls (0.45,1.15) and (0.48,1.65) ..
        (0.52,2.05);

    \draw[boundary]
        (0.18,0.70)
        .. controls (0.39,0.34) and (0.60,0.04) ..
        (0.90,-0.25);

    \draw[boundary]
        (0.90,-0.25)
        .. controls (1.37,-0.20) and (1.80,-0.05) ..
        (2.22,0.15);

    \draw[dashedboundary]
        (0.90,-0.25)
        .. controls (0.58,-0.80) and (0.52,-1.33) ..
        (0.45,-1.95);

    \heavyhalo{(-1.28,1.25)}{halopurple}{0.95}
    \heavyhalo{( 0.90,1.30)}{halopurple}{1.15}
     \heavyhalo{(-0.35,1.62)}{halopurple}{0.55}
     
    \heavyhalo{(-0.95,-1.08)}{halogreen}{1.12}

    \heavyhalo{(-0.38,0.05)}{halogreen}{0.82}
    \heavyhalo{( 0.94,0.82)}{halopurple}{0.92}
    \heavyhalo{( 1.12,-1.05)}{haloorange}{1.02}
    \heavyhalo{( 1.82,-1.00)}{haloorange}{0.78}

    \node[bluepoint]   at (-1.28,1.25) {};
    \node[bluepoint]   at ( 0.90,1.30) {};
    \node[bluepoint] at ( 0.94,0.82) {};

    \node[greenpoint]  at (-0.95,-1.08) {};
    \node[bluepoint]  at (-0.35,1.62) {};

    \node[greenpoint] at (-0.38,0.05) {};
    
    \node[orangepoint] at ( 1.12,-1.05) {};
    \node[orangepoint] at ( 1.82,-1.00) {};

\end{scope}




\end{tikzpicture}
    }

}\subfloat[\footnotesize Symmetry-based sampling]{
    \scalebox{0.75}{
\begin{tikzpicture}[
    x=1cm,
    y=1cm,
    every node/.style={font=\small}
]






\node[
    align=center,
    font=\bfseries
] at (1.53,9.77)
{ Full enumeration};

\node[
    align=center,
    font=\bfseries
] at (5.43,9.77)
{ Sampled representative};

\begin{scope}[shift={(0.68,5.17)}]

    \draw[axis]
        (0,0) -- (2.18,0);

    \draw[axis]
        (0,0) -- (0,3.95);

    \node[below] at (0.22,0) {\Large $x_1$};
    \node[left]  at (0,0.60) {\Large  $x_2$};
    \node[left]  at (0,2.00) {\Large $\vdots$};
    \node[left]  at (0,3.60) {\Large  $x_K$};

    \foreach \i in {0,...,7}{
        \foreach \j in {0,...,11}{
            \fill[gridgray]
                ({0.35+0.22*\i},{0.28+0.29*\j})
                circle (0.055);
        }
    }

    \fill[white]
        (0.22,1.72) rectangle (1.98,1.96);

    \node at (1.10,1.84) {\Large $\cdots$};


\end{scope}

\draw[mainarrow]
    (3.12,7.12) -- (3.93,7.12);


\node[bluepoint] at (4.68,8.78) {};
\node[bluepoint] at (5.26,8.78) {};
\node[anchor=west, font=\Large] at (5.68,8.78) { $\cdots$};

\node[greenpoint] at (4.68,8.02) {};
\node[greenpoint] at (5.26,8.02) {};
\node[anchor=west] at (5.68,8.02) {\Large $\cdots$};

\node[orangepoint] at (4.68,7.26) {};
\node[orangepoint] at (5.26,7.26) {};
\node[anchor=west] at (5.68,7.26) {\Large $\cdots$};

\node[purplepoint] at (4.68,6.50) {};
\node[purplepoint] at (5.26,6.50) {};
\node[anchor=west] at (5.68,6.50) {\Large $\cdots$};

\node[font=\Large] at (5.25,5.78) {\Large $\vdots$};

\node[
    align=center,
    font=\bfseries
] at (5.30,5.00)
{Histogram};


\minihistogram
    {3.30}{3.03}
    {mainblue}
    {0.20/0.38,0.39/0.72,0.58/0.28}

\minihistogram
    {4.30}{3.03}
    {cadmiumgreen}
    {0.20/0.32,0.39/0.62,0.58/0.48}

\minihistogram
    {5.30}{3.03}
    {pointorange}
    {0.20/0.55,0.39/0.76,0.58/0.35}

\node at (6.27,3.54) {\Large $\cdots$};

\minihistogram
    {6.47}{3.03}
    {symbolred}
    {0.20/0.30,0.39/0.69,0.58/0.42}

\fill[
    lightblue,
    rounded corners=5pt
] (0.42,10.58) rectangle (6.93,12.17);

\draw[
    darkblue,
    line width=0.85pt,
    rounded corners=5pt
] (0.42,10.58) rectangle (6.93,12.17);

\node[
    align=center,
    font=\bfseries\large
] at (3.68,11.37)
{reduce offline design complexity\\
by exploiting function symmetry};

\end{tikzpicture}
    
    }

}
\subfloat[\footnotesize MIMO Vector computation]{

\scalebox{0.75}{
\begin{tikzpicture}[
    x=1cm,
    y=1cm,
    every node/.style={font=\small}
]

\node[
    font=\bfseries\footnotesize,
    align=center
] at (1.15,9.78)
{Transmitting nodes};

\node[
    font=\bfseries\footnotesize,
    align=center
] at (5.25,9.78)
{Multi-antenna receiver\\
($M$ antennas)};


\node at (0.15,8.88) {$1$};
\node[blue symbol] at (0.56,8.88) {};
\antenna{1.20}{8.88}

\node at (0.15,7.55) {$2$};
\node[green symbol] at (0.56,7.55) {};
\antenna{1.20}{7.55}

\node[font=\Large] at (0.55,6.66) {$\vdots$};

\node at (0.15,5.75) {$K$};
\node[purple symbol] at (0.56,5.75) {};
\antenna{1.20}{5.75}

\node[mimo block] (mimo) at (3.18,7.34)
{
    \textbf{MIMO}\\[-1mm]
    \textbf{channel}\\[2mm]
    $\mathbf H$
};

\draw[
    flow arrow,
    draw=txblue
]
    (1.48,8.82)
    -- ([yshift=0.78cm]mimo.west);

\draw[
    flow arrow,
    draw=txgreen
]
    (1.48,7.50)
    -- (mimo.west);

\draw[
    flow arrow,
    draw=symbolred
]
    (1.48,5.82)
    -- ([yshift=-0.78cm]mimo.west);

\antenna{4.52}{8.80}
\antenna{4.52}{7.38}
\antenna{4.52}{5.95}


\node[
    anchor=west,
    font=\Large
] at (4.88,7) {$\vdots$};


\draw[
    flow arrow,
    draw=boxpurple
]
    ([yshift=0.72cm]mimo.east)
    -- (4.20,8.75);

\draw[
    flow arrow,
    draw=boxpurple
]
    (mimo.east)
    -- (4.20,7.35);

\draw[
    flow arrow,
    draw=boxpurple
]
    ([yshift=-0.72cm]mimo.east)
    -- (4.20,6.00);

\node[combiner block] (combiner) at (6.18,7.34)
{
    \textbf{Linear}\\[-1mm]
    \textbf{combiner}\\[2mm]
    $\mathbf W$
};

\draw[
    -{Latex[length=2mm,width=1.4mm]},
    line width=0.8pt
]
    (4.83,8.80)
    -- ([yshift=0.78cm]combiner.west);

\draw[
    -{Latex[length=2mm,width=1.4mm]},
    line width=0.8pt
]
    (4.83,7.38)
    -- (combiner.west);

\draw[
    -{Latex[length=2mm,width=1.4mm]},
    line width=0.8pt
]
    (4.83,5.95)
    -- ([yshift=-0.78cm]combiner.west);

\node[
    font=\bfseries\footnotesize,
    align=center
] at (3.23,4.65)
{Vector output / multiple function outputs};

\begin{scope}[shift={(0.30,1.55)}]

    \draw[axis]
        (0,0) -- (3.20,0);

    \draw[axis]
        (0,0) -- (0,2.60);

    \node[right] at (3.12,0) {$I$};
    \node[above] at (0,2.52) {$Q$};

    \draw[
        dashed,
        line width=0.75pt
    ]
        (1.57,0.08)
        .. controls (1.72,0.85) and (1.90,1.68) ..
        (2.05,2.50);

    \draw[
        dashed,
        line width=0.75pt
    ]
        (0.08,1.28)
        .. controls (0.90,1.20) and (2.05,1.45) ..
        (3.08,1.18);

    \foreach \x/\y in {
        0.57/1.83,
        0.79/2.02,
        1.01/1.88,
        0.70/1.65,
        1.11/1.68,
        0.90/1.80,
        0.55/2.10,
        1.22/1.95
    }{
        \fill[txblue] (\x,\y) circle (0.065);
    }

    \foreach \x/\y in {
        2.18/1.90,
        2.43/2.10,
        2.70/1.90,
        2.24/1.65,
        2.57/1.72,
        2.82/2.05,
        2.40/2.28,
        2.73/1.60
    }{
        \fill[txgreen] (\x,\y) circle (0.065);
    }

    \foreach \x/\y in {
        0.52/0.67,
        0.78/0.83,
        1.02/0.68,
        0.63/0.42,
        0.96/0.35,
        1.16/0.55,
        0.48/0.96,
        0.83/0.58
    }{
        \fill[txorange] (\x,\y) circle (0.065);
    }

    \foreach \x/\y in {
        2.17/0.71,
        2.42/0.86,
        2.67/0.70,
        2.27/0.43,
        2.55/0.38,
        2.83/0.56,
        2.39/0.62,
        2.70/0.95
    }{
        \fill[symbolred] (\x,\y) circle (0.065);
    }

\end{scope}

\node[
    anchor=east,
    font=\small
] at (4.25,2.80)
{$\mathbf y=$};

\node[
    font=\small
] at (4.75,2.80)
{
    $\begin{bmatrix}
        y_1\\
        y_2\\
        \vdots\\
        y_L
    \end{bmatrix}$
};

\draw[
    -{Latex[length=2mm,width=1.4mm]},
    draw=txblue,
    line width=0.75pt
]
    (5.14,3.52) -- (5.63,3.52);

\node[
    anchor=west,
    text=txblue
] at (5.70,3.52)
{$f_1(\mathbf{s})$};

\draw[
    -{Latex[length=2mm,width=1.4mm]},
    draw=txgreen,
    line width=0.75pt
]
    (5.14,3.06) -- (5.63,3.06);

\node[
    anchor=west,
    text=txgreen!75!black
] at (5.70,3.06)
{$f_2(\mathbf{s})$};

\node at (5.85,2.61) {$\vdots$};

\draw[
    -{Latex[length=2mm,width=1.4mm]},
    draw=symbolred,
    line width=0.75pt
]
    (5.14,2.12) -- (5.63,2.12);

\node[
    anchor=west,
    text=symbolred
] at (5.70,2.12)
{$f_L(\mathbf{s})$};




\end{tikzpicture}

}
}
\caption{A design toolbox for programming reliable and scalable computation into communication channels. Task- and distribution-aware geometry protects the output distinctions that matter most under the prevailing channel conditions; symmetry reduction and pyramid sampling reduce offline design complexity; and multi-antenna processing enables robust, parallel computation of vector-valued outputs.}
\label{fig:computational-modulation-toolbox}
\end{figure*}
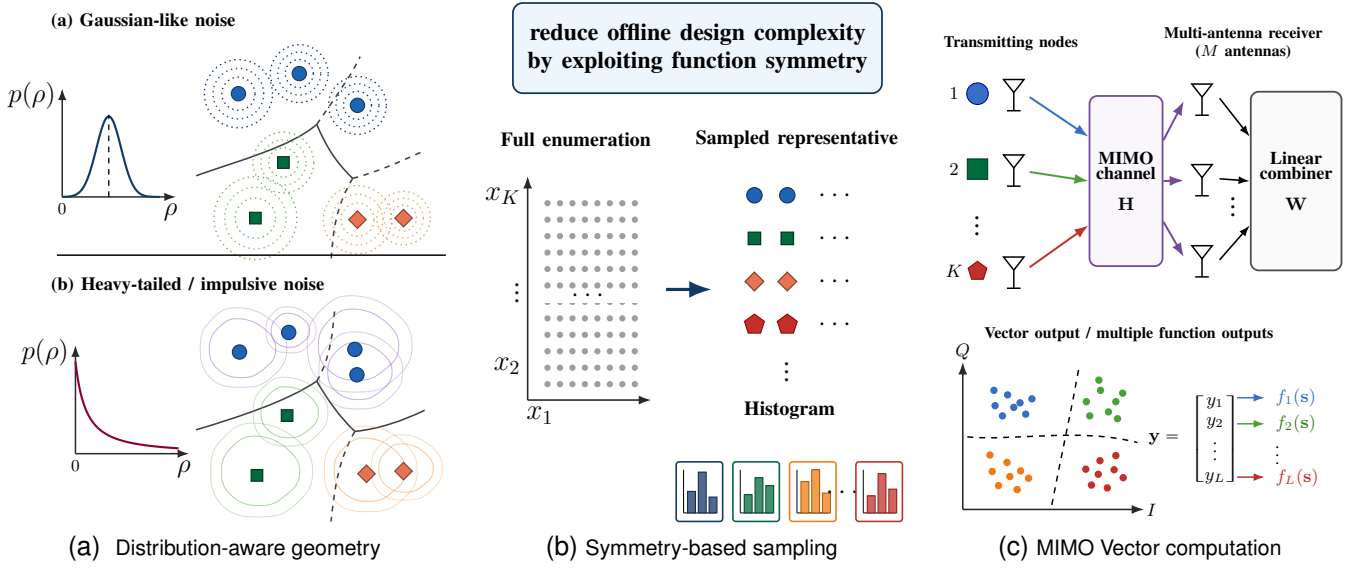

%% file: figures/Fig_SumComp.tex
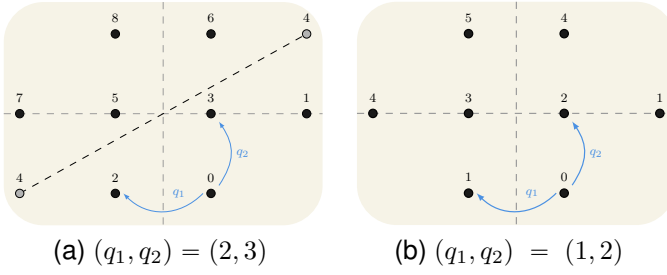
\begin{figure}[!t]
    \centering
    \subfloat[$(q_{1},q_{2}) =(2,3)$]{
    
\scalebox{0.6}{


\begin{tikzpicture}[x=0.75pt,y=0.75pt,yscale=-1,xscale=1]

\draw[draw opacity=0][fill=eggshell!60 , rounded corners=25pt] (100pt, 150pt) rectangle (300pt, 10pt) {};

\draw [color={rgb, 255:red, 155; green, 155; blue, 155 }] [dash pattern={on 4.5pt off 4.5pt}]  (100pt,80pt) -- (300pt,80pt) ;

\draw [color={rgb, 255:red, 155; green, 155; blue, 155 }] [dash pattern={on 4.5pt off 4.5pt}]  (200pt,10pt) -- (200pt,150pt) ;

 \draw[fill=black!90] (110pt,80pt) node{}  circle  (3.5);
 \draw (110pt,70pt) node  {\footnotesize $7$};

\draw[fill=black!90] (170pt,80pt) node{}  circle  (3.5);
\draw (170pt,70pt) node  {\footnotesize $5$};

\draw[fill=black!90] (230pt,80pt) node{}  circle  (3.5);
\draw (230pt,70pt) node  {\footnotesize $3$};

\draw[fill=black!90] (290pt,80pt) node{}  circle  (3.5);
\draw (290pt,70pt) node  {\footnotesize $1$};

\draw[fill=black!90] (170pt,30pt) node{}  circle  (3.5);
\draw (170pt,20pt) node  {\footnotesize $8$};

\draw[fill=black!90] (170pt,130pt) node{}  circle  (3.5);
\draw (170pt,120pt) node  {\footnotesize $2$};

\draw[fill=black!90] (230pt,30pt) node{}  circle  (3.5);
\draw (230pt,20pt) node  {\footnotesize $6$};

\draw[fill=black!90] (230pt,130pt) node{}  circle  (3.5);
\draw (230pt,120pt) node  {\footnotesize $0$};

\draw [color={rgb, 255:red, 74; green, 144; blue, 226 }] (210pt,130pt) node  {\footnotesize $q_1$};

\draw [-latex, color={rgb, 255:red, 74; green, 144; blue, 226 }] (225pt,130pt) .. controls (210pt,145pt) and (190pt,145pt) .. (175pt,130pt) ;

\draw [-latex, color={rgb, 255:red, 74; green, 144; blue, 226 }] (235pt,125pt) .. controls (245pt,110pt) and (245pt,100pt) .. (235pt,85pt) ;

\draw [color={rgb, 255:red, 74; green, 144; blue, 226 }] (250pt,105pt) node  {\footnotesize $q_2$};

\draw  [dash pattern={on 4.5pt off 4.5pt}]  (110pt,130pt) -- (290pt,30pt) ;

\draw[fill=black!30] (110pt,130pt) node{}  circle  (3.5);
\draw (110pt,120pt) node  {\footnotesize $4$};

\draw[fill=black!30] (290pt,30pt) node{}  circle  (3.5);
\draw (290pt,20pt) node  {\footnotesize $4$};

\end{tikzpicture}}
    }\subfloat[$( q_{1} ,q_{2}) \ =\ ( 1,2)$]{

\tikzset{every picture/.style={line width=0.75pt}} 
\scalebox{0.6}{
\begin{tikzpicture}[x=0.75pt,y=0.75pt,yscale=-1,xscale=1]

\draw[draw opacity=0][fill=eggshell!60 , rounded corners=25pt] (100pt, 150pt) rectangle (300pt, 10pt) {};

\draw [color={rgb, 255:red, 155; green, 155; blue, 155 }] [dash pattern={on 4.5pt off 4.5pt}]  (100pt,80pt) -- (300pt,80pt) ;

\draw [color={rgb, 255:red, 155; green, 155; blue, 155 }] [dash pattern={on 4.5pt off 4.5pt}]  (200pt,10pt) -- (200pt,150pt) ;

 \draw[fill=black!90] (110pt,80pt) node{}  circle  (3.5);
 \draw (110pt,70pt) node  {\footnotesize $4$};

\draw[fill=black!90] (170pt,80pt) node{}  circle  (3.5);
\draw (170pt,70pt) node  {\footnotesize $3$};

\draw[fill=black!90] (230pt,80pt) node{}  circle  (3.5);
\draw (230pt,70pt) node  {\footnotesize $2$};

\draw[fill=black!90] (290pt,80pt) node{}  circle  (3.5);
\draw (290pt,70pt) node  {\footnotesize $1$};

\draw[fill=black!90] (170pt,30pt) node{}  circle  (3.5);
\draw (170pt,20pt) node  {\footnotesize $5$};

\draw[fill=black!90] (170pt,130pt) node{}  circle  (3.5);
\draw (170pt,120pt) node  {\footnotesize $1$};

\draw[fill=black!90] (230pt,30pt) node{}  circle  (3.5);
\draw (230pt,20pt) node  {\footnotesize $4$};

\draw[fill=black!90] (230pt,130pt) node{}  circle  (3.5);
\draw (230pt,120pt) node  {\footnotesize $0$};

\draw [color={rgb, 255:red, 74; green, 144; blue, 226 }] (210pt,130pt) node  {\footnotesize $q_1$};

\draw [-latex, color={rgb, 255:red, 74; green, 144; blue, 226 }] (225pt,130pt) .. controls (210pt,145pt) and (190pt,145pt) .. (175pt,130pt) ;

\draw [-latex, color={rgb, 255:red, 74; green, 144; blue, 226 }] (235pt,125pt) .. controls (245pt,110pt) and (245pt,100pt) .. (235pt,85pt) ;

\draw [color={rgb, 255:red, 74; green, 144; blue, 226 }] (250pt,105pt) node  {\footnotesize $q_2$};

\end{tikzpicture}}
    }
    \caption{Programming an order-8 hexagonal-QAM constellation for summation. The parameters $(q_1,q_2)$ determine how integer values are distributed across the two lattice directions. The two configurations support different input ranges and label repetitions while preserving arithmetic addition after superposition; arbitrary relabeling can instead produce ambiguous aggregate points~\cite{Razavikia2025SumComp}.}
    \label{fig:Hexagonal}
\end{figure}

%% file: figures/Table_source.tex
\begin{table}[!t]
\caption{Uplink Resource Accounting}
\label{tab:feel-resource-accounting}
\centering
\footnotesize
\setlength{\tabcolsep}{3pt}
\renewcommand{\arraystretch}{1.12}
\begin{tabular}{
    >{\raggedright\arraybackslash}p{0.65\columnwidth}
    >{\centering\arraybackslash}p{0.25\columnwidth}}
\hline
\rowcolor{mainblue!20} 
\textbf{Uplink scheme} &
\textbf{Data REs} \\
\hline
\rowcolor{lightblue!75}
Digital QAM aggregation &
\(N\) \\
Uncompressed analog OAC &
\(N\) \\
\rowcolor{lightblue!75}
 Orthogonal digital upload &
\(KN\) \\
\hline
\end{tabular}

\vspace{2pt}
\parbox{\columnwidth}{\scriptsize
Common assumptions: \(K\) devices, \(N\) coordinates per update, one coordinate represented per transmitted symbol, one spatial stream, and no compression or channel coding. The counts include only data-bearing time--frequency resource elements; pilots, guard intervals, control signaling, and retransmissions are excluded.}
\end{table}

%% file: figures/Fig_fed.tex
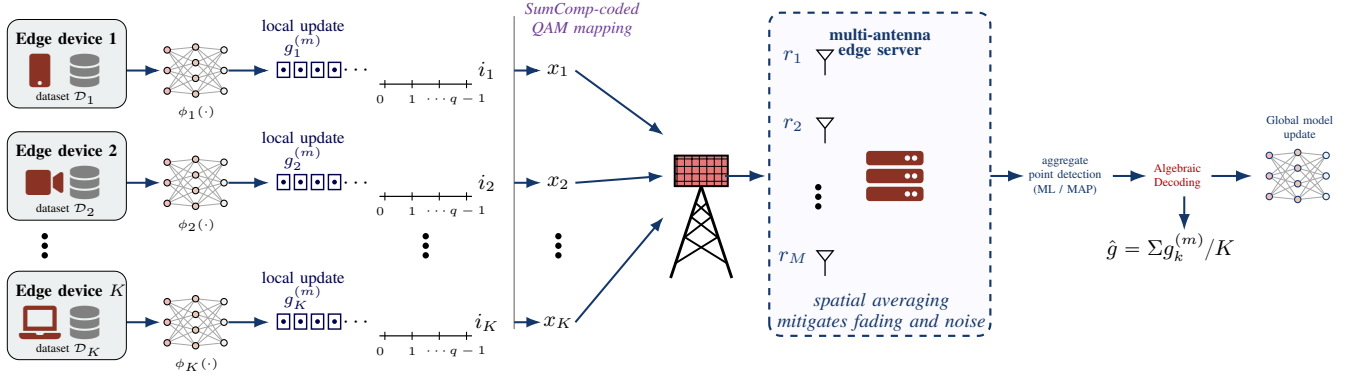
\begin{figure*}
    \centering
    \scalebox{0.9}{

\begin{tikzpicture}[
    x=1cm,
    y=1cm,
    every node/.style={font=\small}
]


\def\yA{4.85}
\def\yB{3.20}
\def\yC{1.15}

\draw[rowflow]
    (1.72,\yA)--(2.25,\yA);

\draw[rowflow]
    (1.72,\yB)--(2.25,\yB);

\draw[rowflow]
    (1.72,\yC)--(2.25,\yC);


\draw[fill=palecgray , rounded corners=5pt] (0, \yA-0.55) rectangle (1.75, \yA+0.75) {};

\node[
    anchor=west,
    font=\scriptsize\bfseries
] at (0,5.3)
{Edge device 1};


\node[anchor=west]
    at (0.2,\yA) {{\color{burntumber} \Large \faMobile}};


\node[anchor=west,font=\tiny]
    at (0.3,\yA-0.4) {dataset $\mathcal D_1$};


\node[anchor=west]
    at (0.8,\yA) {{\color{gray} \Large \faDatabase}};
    

\draw[fill=palecgray , rounded corners=5pt] (0, \yB-0.55) rectangle (1.75, \yB+0.75) {};

\node[
    anchor=west,
    font=\scriptsize\bfseries
] at (0,3.65)
{Edge device 2};


\node[anchor=west]
    at (0.15,\yB) {{\color{burntumber} \Large \faVideo}};

\node[anchor=west]
    at (0.8,\yB) {{\color{gray} \Large \faDatabase}};


\node[anchor=west,font=\tiny]
    at (0.3,\yB-0.4) {dataset $\mathcal D_2$};


\node[font=\Huge] at (0.55,2.38) {$\vdots$};


\draw[fill=palecgray , rounded corners=5pt] (0, \yC-0.55) rectangle (1.75, \yC+0.75) {};

\node[
    anchor=west,
    font=\scriptsize\bfseries
] at (0,1.6)
{Edge device $K$};

\node[anchor=west]
    at (0.05,\yC) {{\color{burntumber} \Large \faLaptop}};

\node[anchor=west]
    at (0.8,\yC) {{\color{gray} \Large \faDatabase}};

\node[anchor=west,font=\tiny]
    at (0.3,\yC-0.4) {dataset $\mathcal D_K$};


\nnicon{2.35}{\yA}{black}
\nnicon{2.35}{\yB}{black}
\nnicon{2.35}{\yC}{black}

\node[font=\tiny] at (2.78,\yA-0.62)
    {$\phi_1(\cdot)$};

\node[font=\tiny] at (2.78,\yB-0.62)
    {$\phi_2(\cdot)$};

\node[font=\tiny] at (2.78,\yC-0.62)
    {$\phi_K(\cdot)$};

\draw[rowflow]
    (3.25,\yA)--(3.85,\yA);

\draw[rowflow]
    (3.25,\yB)--(3.85,\yB);

\draw[rowflow]
    (3.25,\yC)--(3.85,\yC);

\localupdate{4.35}{\yA}{1}{blue!30!black}
\localupdate{4.35}{\yB}{2}{blue!30!black}
\localupdate{4.35}{\yC}{K}{blue!30!black}

\node at (5.15,\yA) {$\cdots$};
\node at (5.15,\yB) {$\cdots$};
\node at (5.15,\yC) {$\cdots$};

\quantizer{6.15}{\yA}
\quantizer{6.15}{\yB}
\quantizer{6.15}{\yC}

\node[font=\Huge] at (6.15,2.38) {$\vdots$};

\node at (7.05,\yA) {$i_1$};
\node at (7.05,\yB) {$i_2$};
\node at (7.05,\yC) {$i_K$};

\draw[black!60,line width=0.45pt]
    (7.45,1.05)--(7.45,5.65);

\node[
    align=center,
    text=violet,
    font=\scriptsize\itshape
] at (8.45,5.60)
{SumComp-coded\\QAM mapping};

\draw[rowflow]
    (7.45,\yA)--(7.85,\yA);

\draw[rowflow]
    (7.45,\yB)--(7.85,\yB);

\draw[rowflow]
    (7.45,\yC)--(7.85,\yC);

\node at (8.10,\yA) {$x_1$};
\node at (8.10,\yB) {$x_2$};
\node at (8.10,\yC) {$x_K$};

\node[font=\Huge] at (8.10,2.38) {$\vdots$};


\draw[rowflow=scatterpurple]
    (8.35,\yA)--(9.65,3.90);

\draw[rowflow=scatterpurple]
    (8.35,\yB)--(9.65,3.30);

\draw[rowflow=scatterpurple]
    (8.35,\yC)--(9.65,2.70);


\begin{scope}[scale=0.03, shift={(215cm,195cm)}, yscale=-1]






          \draw[fill=bazaar] (112,75) rectangle (138,90);
   \draw[step=3.65,black,thin] (112,75) grid (138,90);
  \draw[black,thick]   (112,75) rectangle (138,90);

\draw[very thick]    (110,135) -- (125,90) ;
\draw[very thick]    (140,135) -- (125,90) ;

\draw[thick]    (112,128) -- (133,115) ;
\draw[thick]    (138,128) -- (117,115) ;

\draw[thick]    (133,115) --  (120,105)  ;
\draw[thick]    (117,115)  -- (130,105)  ;

\draw[thick]     (120,105) -- (128,98) ;
\draw[thick]     (130,105)  -- (122,98) ;
\end{scope}

\draw[flow]
    (10.55,3.30)--(11.15,3.30);

\draw[receiver]
    (11.20,0.95) rectangle (14.45,5.70);

\node[
    text=mainblue,
    font=\scriptsize\bfseries
] at (12.83,5.38)
{multi-antenna};

\node[
    text=mainblue,
    font=\scriptsize\bfseries
] at (12.83,5.13)
{edge server};

\node[text=mainblue] at (11.55,5.05) {$r_1$};
\rxant{12.02}{5.00}

\node[text=mainblue] at (11.55,4.05) {$r_2$};
\rxant{12.02}{4.00}

\node[font=\Huge] at (11.92,3.10) {$ \vdots$};

\node[text=mainblue] at (11.55,2.10) {$r_M$};
\rxant{12.02}{2.05}


\node[anchor=west, font=\Huge]
    at (12.5,3.3) {{\color{burntumber} \faServer}};




\node[
    align=center,
    text=mainblue,
    font=\footnotesize\itshape
] at (12.85,1.30)
{spatial averaging\\mitigates fading and noise};

\draw[flow]
    (14.45,3.30)--(14.95,3.30);

\node[
    align=center,
    text=mainblue,
    font=\tiny
] at (15.55,3.30)
{aggregate\\point detection\\(ML / MAP)};


\node[
    align=center,
    text=red!70!black,
    font=\tiny
] at (17.2,3.3)
{Algebraic\\Decoding};

\draw[flow]
    (16.25,3.30)--(16.75,3.30);


\draw[flow]
    (17.7,3.30)--(18.3,3.30);

\node at (17.1,2.25)
    {$\hat{g}= {\Sigma} g_k^{(m)}/K$};

\draw[flow]
    (17.3,3)--(17.3,2.5);

\node[
    align=center,
    text=mainblue,
    font=\tiny
] at (19,4)
{Global model\\update};

\nnicon{18.55}{3.30}{mainblue}








\end{tikzpicture}

    }
    \caption{Programming federated aggregation into a shared uplink. Each device quantizes its local model update, maps it to SumComp-coded QAM symbols, and transmits simultaneously. The multi-antenna edge server mitigates fading and noise, detects the aggregate constellation point, and algebraically recovers the average update without reconstructing the individual contributions.}
    \label{fig:digital-computational-channel}
\end{figure*}

%% file: figures/Table_compare.tex
\begin{table*}[t]
\caption{Comparison of Representative Over-the-Air Computation Approaches}
\label{tab:method-synthesis}
\centering
\scriptsize
\setlength{\tabcolsep}{4pt}
\renewcommand{\arraystretch}{1.2} 
\begin{tabular}{
>{\raggedright\arraybackslash}p{2.1cm}
>{\raggedright\arraybackslash}p{3.0cm}
>{\raggedright\arraybackslash}p{3.1cm} 
>{\raggedright\arraybackslash}p{2.1cm}
>{\raggedright\arraybackslash}p{2.0cm}
>{\raggedright\arraybackslash}p{3.5cm}}
\toprule

\rowcolor{mainblue!20} 
\textbf{Approach} &
\textbf{Function and channel} &
\textbf{CSI and synchronization} &
\textbf{Complexity} &
\textbf{Channel uses\textsuperscript{*}} &
\textbf{Main trade-off} \\
\midrule

\rowcolor{lightblue!75} 
Analog OAC &
Sums over a fading MAC &
Tx CSI and tight alignment &
Low design; continuous adaptation &
One per scalar sum &
Fast aggregation, but sensitive to channel errors \\
\addlinespace 

ChannelComp~\cite{Razavikia2024ChannelComp} &
Finite-alphabet functions over an additive MAC &
Model dependent; coherent reception &
High design; simple operation &
One per scalar output &
Supports many functions, but design scales poorly \\
\addlinespace

\rowcolor{lightblue!75}
SumComp~\cite{Razavikia2025SumComp} &
Integer sums using PAM or QAM &
Mapping: CSI-independent; fading: Tx inversion or blind Rx MIMO compensation~\cite{razavikia2024blind} &
Closed-form and low &
One per scalar sum &
Simple implementation, but mainly supports sums \\
\addlinespace

VecComp~\cite{Razavikia2026VecComp} &
Vector functions over a MIMO MAC &
Receiver CSI; no instantaneous Tx CSI &
Component-wise design &
Multiple outputs per use &
Parallel computation, but requires spatial modes \\
\bottomrule
\end{tabular}

\vspace{4pt} 
\raggedright
\textsuperscript{*} The channel-use counts exclude pilots, coding, and retransmissions.
\end{table*}